# Stabilization of lifted hydrogen jet diffusion flame in a vitiated co-flow: effects of jet and coflow velocities, coflow temperature and mixing

**Santanu De[1], Ashoke De[2]\*, Abhishek Jaiswal[2], Arpita Dash[2]**

[1]Department of Mechanical Engineering, Indian Institute of Technology, Kanpur, India-208016
[2]Department of Aerospace Engineering, Indian Institute of Technology, Kanpur, India-208016

**Abstract:** The present paper reports on the numerical investigation of lifted turbulent jet flames with $H_2/N_2$ fuel issuing into a vitiated coflow of lean combustion products of $H_2/air$ using conditional moment closure method (CMC). A 2D axisymmetric formulation has been used for the predictions of fluid flow, while CMC equations are solved with detailed chemistry to represent the turbulence-chemistry interaction. Simulations are carried out for different coflow temperatures, jet and coflow velocities in order to investigate the impact on the flame lift-off height as well as on the flame stabilization. Furthermore, the role of conditional velocity models on the flame has also been investigated. In addition, the effect of mixing is investigated over a range of coflow temperatures and the stabilization mechanism is determined from the analysis of the transport budgets. It is found that the lift-off height is highly sensitive to the coflow temperature, while the predicted lift-off height using the mixing model constant, *i.e.*, $C_\phi$=4, is found to be the closest to the experimental results. For all the coflow temperatures, the balance is found between the chemical, axial convection and molecular diffusion terms while the contribution from axial and radial diffusion is negligible, thus indicating auto-ignition as the flame stabilization mechanism.

\*Corresponding Author: Tel.: +91-51-2597863, Fax: +91-512-2597561
E-mail address: ashoke@iitk.ac.in

## 1. Introduction

Lifted turbulent jet diffusion flame has been a subject of active research as they exhibit many complex physical phenomena, such as ignition and extinction, which has relevance in many industrial applications [1]. A substantial number of experimental and numerical studies have been carried out to understand and analyze different flame stabilization mechanism in lifted flames, leading to competing theories on flame stabilization. Early experiments on lifted flame concluded that the fuel and oxidizer streams get completely premixed at the base of the flame and flame stabilizes through a balance of local mean velocity and the turbulent flame speed at a location where the mixture becomes stoichiometric [2]. Downstream of the liftoff region resembles to a diffusion flame. This led to premixed flame propagation theory of turbulent lifted flames; and the correlation between lift-off heights and burning velocity in conjunction with other flame properties, *e.g.*, density and viscosity were proposed by Kalghatgi [3]. Muniz and Mungal [4] observed that the flame stabilizes in regions where the local incoming velocity of the reactants is near the burning velocity of a laminar premixed flame. Propagation of triple flames at the flame base has also been believed as a possible stabilization mechanism [5-6]. In a direct numerical simulation (DNS) study of lifted $H_2$ jet diffusion flame in still air [7], also a triple flame like structure was reported. However, Peter and Williams perceived that the level of premixedness is insufficient to sustain a premixed flame at the base of the lifted flame, and argued that the higher level of the mean scalar dissipation rate at the flame base is indeed responsible for the flame stabilization through local quenching of the laminar diffusion flamelets [8]. Later, Schefer *et al.* [9] contradicted the findings of Peter and Wiliams [8] through their experimental measurements, where they found scalar dissipation rate at the flame base is far less than the quenching value at the flame stabilization points. Moreover, the experiments by Tacke *et al.* [10] revealed that the lifted

H$_2$-flame stabilizes in lean mixtures, rather than the stoichiometric mixtures and these results are in complete contradiction with the existing theory available in literatures [2, 4].

The above-mentioned flames are primarily diffusion flames where fuel jets are issued into a surrounding air at the ambient condition. While moving towards the jet flame supported by a hot coflow, Cabra *et al.* [11] conducted the first experimental measurements for a lifted turbulent H$_2$/N$_2$ jet flame supported by an annular, vitiated coflowing oxidizer stream consisting of combustion products of a lean premixed H$_2$/air flame. The inclusion of vitiated coflow allows another interesting possibility of flame stabilization by means of auto-ignition, as the inert fuel parcels emanating from the jet are advected downstream, and get mixed with the surrounding hot coflowing oxidizer stream leading to autoignition of the fuel/air mixture. The Cabra burner [11] has been widely used for development and validation of turbulent combustion models as it offers a simple flow configuration with well-defined inlet boundary conditions and availability of experimental measurements for several key quantities at several downstream locations. This flame has been extensively investigated using numerical techniques by different researchers for varying parameters like coflow temperature, jet velocity and coflow velocity [12-22]. The major findings of the PDF methods were that the flame is primarily kinetically controlled and hence stabilized due to auto-ignition only. However, Gordon *et al.* [16] distinguished between two different flame stabilization mechanisms, *i.e.* auto-ignition and premixed flame propagation, by using numerical indicators based on budget of the convection, diffusion and production of key radical species such as OH. More recently, Yadav *et al.* [18] modelled this flame by invoking presumed shape multi-environment Eulerian PDF (MEPDF) transport method in conjunction with interaction-by-exchange with mean (IEM) model for micro-mixing. However, despite using advanced turbulence-chemistry interaction model, their lift-off height predictions are not in good agreement with measurements. Amongst the other PDF based methods, Wang and Pope [19] investigated the phenomenon of local extinction, re-ignition and auto-ignition of these flames using different mixing models. The impact of burnt gases on flame stabilization has been studied using LES in conjunction with auto-ignition and premixed flamelet chemistry tabulations [20]. In some of the recent DNS works of a $H_2/N_2$ flame in a vitiated coflow [21-23], the flame stabilization mechanism has been investigated in a comprehensive manner. Yoo et al. [21] reported that the mean axial velocity is about an order of magnitude higher than the laminar flame velocity at the stabilization point, which rules out premixed flame propagation as the phenomenon behind the stabilization of the lifted turbulent flame. They conjectured that the flame stabilization is inherently due to a competition of autoignition and large scale structures. A comprehensive review on different stabilization mechanisms for lifted jet flames is available in recent review articles [1, 24-25].

Besides the PDF method, over the last decade, the conditional moment closure (CMC) [26] method has emerged as an alternate model for turbulence-chemistry interaction. It removes the major source of non-linearity from the reaction rate terms by considering the various moments of species concentrations, conditionally averaged at a fixed value of a conserved scalar, namely mixture fraction for non-premixed combustion case, and evaluating the conditional reaction rate terms using the computed conditional mean values of temperature and species concentrations. The CMC method has been successfully applied to a variety of turbulent non-premixed flame configurations, such as jet flames [27], piloted jet diffusion flames [28], bluff-body stabilized flames [29], auto-ignition of fuel jets [30-31] and diluted hot coflow [32]. Solution of the first-order, radially-averaged CMC [33] equations along with the *k-ε* turbulence model predicted lift-off height and flame structure which agreed well with the experimental data. Also, the CMC method yielded the flame structures reasonably well in the moderate and intense low oxygen dilution combustion mode [32]. The predicted conditional mean temperature, CO, OH and NO mass fractions are found to be in good agreement with the experimental measurements. Cabra burner has also been investigated following both Reynolds-averaged Navier Stokes (RANS) [34-35] and large eddy simulation (LES) [36-37] based CMC methods. Patwardhan *et al*. [34] applied CMC method on $H_2/N_2$ flame for different physical parameters and observed both the modes of flame stabilization at different conditions. More recently, El Sayed and Fraser [35] applied CMC to investigate a lifted $H_2/N_2$ Cabra flame using a modified $k$-$\varepsilon$ model and assessed performances of the homogeneous model of Girimaji [38] and the inhomogeneous model of Mortensen [39] for the conditional scalar dissipation rate (CSDR) and the PDF-gradient diffusion model for modeling the conditional velocity. The Cabra burner has also been investigated using LES based CMC method following a first-order closure for the conditional chemical source term [37]. It is concluded that the flame stabilization is controlled by autoignition for all coflow temperatures. Again, adopting the similar methodology as outlined in Ref. [37], Navarro-Martinez and Kronenburg [36] drew similar conclusions. They claimed that at high coflow temperatures, flame stabilization mechanism is controlled by autoignition, however they reported a transition in flame stabilization mechanism at low coflow temperatures, where premixed flame propagation governs the flame stabilization.

However, none of the above-mentioned literature on CMC simulation of a lifted flame in vitiated coflow [34-37] provides a comprehensive detail on the prediction of lift-off height and flame stabilization mechanism due to various physical parameters such as mixing model constant, conditional velocity models, turbulence parameters etc. In CMC, appropriate modeling of conditional velocity plays a key role for improving the accuracy of flow field and scalar predictions. However, no such comparison is available in the literature on different conditional velocity models for lifted jet diffusion flames supported by a vitiated coflow. The sensitivity of the lift-off height to the mixing model constant $C_\Phi$ has been examined in the past using PDF methods [14] and it is found that the lift-off is sensitive to mixing model constant $C_\phi$, and they have proposed different values for different micro-mixing models. However, the effect of $C_\phi$ on the lift-off height has not been investigated within the framework of CMC method.

The present work is directed toward developing an improved fundamental understanding of flame stabilization mechanism which may be useful for industrial applications, *e.g.* combustor design. The primary aim of this work is to investigate and shed light on the understanding of the flame characteristics due to the changes in various physical parameters and modeling strategies such as: (1) effect of conditional velocity sub-models (mean, linear and gradient) on scalar predictions and liftoff height predictions, (2) effect of change in co-flow temperature, co-flow and jet velocities on flame lift-off height, (3) sensitivity of the lift-off height on mixing model constants ($C_\phi$) at different co-flow temperatures and (4) study the fundamental mechanism of flame stabilization at different coflow temperatures, jet and coflow velocities, and using different mixing model constants.

## 2. Mathematical Formulation

In CMC, the reactive scalars such as the species mass fractions and the temperature are conditionally averaged with respect to the mixture fraction, $\xi$, and their conditional transport equations are solved. The conditional average of mass fraction of a species $\alpha$ is defined as,

$$Q_\alpha(\eta, x, t) = \langle \rho Y_\alpha(x,t) | \xi(x,t) = \eta \rangle / \rho_\eta = \langle Y_\alpha | \eta \rangle, \tag{1}$$

where $\xi$ is the mixture fraction and $\eta$ is the sample space variable of $\xi$. The instantaneous scalar quantity $Y_\alpha$ can be decomposed into its Favre conditional mean $Q_\alpha$ and the conditional fluctuation $Y_\alpha''$ as:

$$Y_\alpha(x,t) = Q_\alpha(\eta, x, t) + Y_\alpha''(x,t). \tag{2}$$

Then, the conservation equation for the conditional mean species mass fraction $Q_\alpha$ is given by:

$$\frac{\partial Q_\alpha}{\partial t} = -\langle u|\eta\rangle . \nabla Q_\alpha - \frac{\nabla . \rho_\eta \langle u''Y''_\alpha|\eta P(\eta)\rangle}{\rho_\eta P(\eta)} + \langle N|\eta\rangle \frac{\partial^2 Q_\alpha}{\partial \eta^2} + \frac{1}{\rho_\eta}\langle \dot{\omega}_\alpha|\eta\rangle , \qquad (3)$$

where a high Reynolds number is assumed. Also, the Lewis numbers for all species are assumed to be unity. The first and second terms on the right hand side of Eq.(3) represent transport by means of advection and diffusion in the physical space, the third term accounts for molecular diffusion in mixture fraction space, and the fourth term is the chemical source term. The unconditional or Favre averages $(\widetilde{\phi})$ of the reactive scalars are recovered by convoluting the conditional mean quantities $\langle \phi|\eta\rangle$ with the PDF $\tilde{P}(\eta)$, over the mixture fraction space:

$$\widetilde{\phi} = \int_0^1 \langle \phi|\eta\rangle \tilde{P}(\eta)d\eta, \qquad (4)$$

where, $\phi = \{Y_\alpha, T\}$. The Favre PDF $\tilde{P}(\eta)$ is approximated using the standard $\beta$-function represented in terms of the Favre mean and variance of mixture fraction $(\tilde{\xi}, \widetilde{\xi''})$. In Eq.(3), the quantities $\tilde{P}(\eta), \langle u''Y''_\alpha|\eta\rangle, \langle u|\eta\rangle, \langle N|\eta\rangle$ and $\langle \dot{\omega}_\alpha|\eta\rangle$ are unclosed and require modeling. The conditional turbulent fluxes are modelled by gradient diffusion assumption:

$$\langle u''Y''_\alpha|\eta\rangle = -D_t \frac{\partial Q_\alpha}{\partial x_i}, \qquad (5)$$

where, the turbulent diffusivity is:

$$D_t = \frac{C_\mu}{Sc_t}\frac{\tilde{k}^2}{\tilde{\varepsilon}}. \qquad (6)$$

The constant $C_\mu$ and the turbulent Schmidt number ($Sc_t$) are taken as 0.09 and 0.7, respectively. In the present study, the conditional velocity is approximated using the following three models:

Mean: $\qquad \langle u_i|\eta\rangle = \tilde{u}_i, \qquad (7)$

Linear: $\qquad \langle u_i|\eta\rangle = \tilde{u}_i + \frac{\widetilde{u''_i \xi''}}{\widetilde{\xi''^2}}(\eta - \tilde{\xi}), \qquad (8)$

Gradient diffusion: $\quad \langle u_i | \eta \rangle = \tilde{u}_i - \frac{D_t}{\tilde{P}(\eta)} \frac{\partial \tilde{P}(\eta)}{\partial x_i}.$ \hfill (9)

The gradient diffusion model when used with the presumed PDF methods is entirely consistent with the moments of conserved scalars.

In the first-order CMC, conditional fluctuation of any gas-phase reactive scalars about its conditional mean is assumed to be small, and therefore the conditional mean chemical source term $\langle \omega_\alpha | \eta \rangle$ is modelled as a function of the conditional mean temperature and species mass fractions:

$$\langle \dot{\omega}_\alpha (\rho, Y_1, Y_2, \dots, Y_{N_S}, T | \eta \rangle = \dot{\omega}_\alpha (\rho_\eta, Q_1, Q_2, \dots, Q_{N_S}, Q_T), \text{ where } \alpha = 1,2,\dots N_s. \quad (10)$$

The conditional mean of the scalar dissipation rate is modeled by the amplitude mapping closure method [40]:

$$\langle N | \eta \rangle = \frac{\tilde{\chi}}{2} \frac{G(\eta)}{\int_0^1 G(\eta) P(\eta) d\eta}, \hfill (11)$$

$$G(\eta) = \exp\{-2[erf^{-1}(2\eta - 1)]^2\}. \hfill (12)$$

## 2.1 Turbulent flow field calculation

Favre-averaged Navier-Stokes equations are solved in order to obtain the velocity $(\tilde{u})$ and pressure $(\tilde{p})$ field. The turbulent flow field is modeled using the $k$-$\varepsilon$ model. The transport equations of the turbulent kinetic energy $(\tilde{k})$ and its dissipation rate $(\tilde{\varepsilon})$ are given by:

$$\bar{\rho} \frac{\partial \tilde{k}}{\partial t} + \bar{\rho} \tilde{u}_j \frac{\partial \tilde{k}}{\partial x_j} = \frac{\partial}{\partial x_j} \left[ \left( \frac{\mu_t}{\sigma_k} + \mu \right) \frac{\partial \tilde{k}}{\partial x_j} \right] - \overline{\rho u_i'' u_j''} \frac{\partial \tilde{u}_i}{\partial x_j} - \rho \tilde{\varepsilon} , \hfill (13)$$

$$\bar{\rho} \frac{\partial \tilde{\varepsilon}}{\partial t} + \bar{\rho} \tilde{u}_j \frac{\partial \tilde{\varepsilon}}{\partial x_j} = \frac{\partial}{\partial x_j} \left[ \left( \frac{\mu_t}{\sigma_\varepsilon} + \mu \right) \frac{\partial \tilde{\varepsilon}}{\partial x_j} \right] - C_{\varepsilon 1} \frac{\tilde{\varepsilon}}{\tilde{k}} \left( \overline{\rho u_i'' u_j''} \frac{\partial \tilde{u}_i}{\partial x_j} \right) - C_{\varepsilon 2} \bar{\rho} \frac{\widetilde{\varepsilon^2}}{\tilde{k}} , \hfill (14)$$

where, $\sigma_k$ and $\sigma_\varepsilon$ are the turbulent Prandtl/Schmidt number for $\tilde{k}$ and $\tilde{\varepsilon}$ respectively, $C_{\varepsilon 1}$ and $C_{\varepsilon 2}$ are the constants used in the $k$-$\varepsilon$ model, and,

$$\mu_t = \bar{\rho} C_\mu \frac{\widetilde{k^2}}{\widetilde{\varepsilon}}$$ 

( 15 )

is the turbulent viscosity. In this study, the default model constants, i.e. $C_{\varepsilon 1} = 1.44$ and $C_{\varepsilon 2} = 1.92$, have been used after carrying out a systematic studies.

## 2.2 Mixing field calculation

In case of non-premixed combustion and under assumption of equal diffusivities and unity Lewis number, the several dependent variables $Y_\alpha$ can be represented by a single conserved scalar, *i.e.*, mixture fraction ($\xi$) [41]. The transport equation for the mean mixture fraction ($\tilde{\xi}$) is represented by,

$$\bar{\rho} \frac{\partial \tilde{\xi}}{\partial t} + \bar{\rho} \widetilde{u_j} \frac{\partial \tilde{\xi}}{\partial x_j} = \frac{\partial}{\partial x_j} \left( \bar{\rho} D \frac{\partial \tilde{\xi}}{\partial x_j} - \overline{\rho u_j'' \xi''} \right).$$

( 16 )

There are no source terms in the above equation and it describes the mixing of gaseous fuel and oxidizer in a turbulent flow-field. In addition to mean mixture fraction transport, the following transport equation for variance of mixture fraction ($\widetilde{\xi''^2}$), is solved,

$$\bar{\rho} \frac{\partial \widetilde{\xi''^2}}{\partial t} + \bar{\rho} \widetilde{u_j} \frac{\partial \widetilde{\xi''^2}}{\partial x_j} = -\frac{\partial}{\partial x_j} \left( \bar{\rho} \widetilde{u_j'' \xi''^2} \right) - 2 \bar{\rho} \widetilde{u_j'' \xi''} \frac{\partial \tilde{\xi}}{\partial x_j} - \bar{\rho} \tilde{\chi},$$

( 17 )

where, the Favre mean (average) scalar dissipation rate $\tilde{\chi}$ is modeled using the traditional assumption of equality of the time scales [26-35] as:

$$\tilde{\chi} = C_\phi \widetilde{\xi''^2} \left( \frac{\widetilde{\varepsilon}}{\widetilde{k}} \right).$$

( 18 )

The constant $C_\Phi$ represents the ratio of the turbulent and mixing time scales. The influence of change of $C_\Phi$ on lift-off height has been discussed in Sec.4.6.

## 3. Numerical Methods

In this study, the lifted turbulent $H_2/N_2$ jet diffusion flame is modelled following the burner configuration of Cabra *et al.* [11]. A schematic of the experimental setup is shown in Figure 1. The central jet has an outer diameter (*D*) of 4.57 mm which is surrounded by a co-flow of lean combustion products of H$_2$. The boundary conditions used in the present numerical simulations are given in Table 1. Further details on the burner configuration and experimental methods are available in the literature by Cabra *et al.* [11].

The mean flow is assumed to be axisymmetric, and the governing equations in a cylindrical coordinate system [34] are solved. The computational domain extends to 30$D$ and 12.5$D$ in the axial- and radial directions, respectively. The numerical simulations are carried out using three different sets of grids: (a) a coarse grid with 72X72 cells, (b) an intermediate grid with 96X96 cells, and (c) a fine grid with 120X120 cells in the radial- and axial directions, respectively. The mesh is refined near the centerline and within the shear layer. The smallest mesh sizes in the axial and radial directions are 1.48 mm and 0.0863 mm for the intermediate grid. The governing equations are solved using a pressure-based, fully-elliptic, finite volume formulation in cylindrical coordinate system. Spatial diffusion terms appearing in the governing equations are discretized following a second-order central-difference scheme, whereas advective terms are discretized using an upwind-difference scheme with the Koren limiter ($\kappa$=1/3). Time integration is performed following the fractional step method using the second-order Adams-Bashforth scheme for the convection and diffusion terms in physical space and the Crank-Nicolson method for diffusion term in the mixture fraction space and chemical source term. The numerical code used in the present work, was previously used in the RANS-CMC modeling for lifted jet diffusion flames [34] and autoignition of turbulent hydrogen jet in a hot coflow [31]. In order to solve the CMC equations, the mixture fraction space is discretized into 62 cells of variable width clustered around $\eta$=0, $\eta_{st}$, and 1. At every time step, the flow and mixing field variables ($\tilde{u}, \overline{p}, \tilde{k}, \tilde{\varepsilon}, \tilde{\xi}, \widetilde{\xi''}$) obtained from the turbulent flow calculations are passed to the CMC solver, where the transport equations for the conditional scalars ($Q_\alpha, Q_T$) are solved in the mixture fraction space. The Favre mean density field, obtained by convoluting the conditional density with $\tilde{P}(\eta)$, is sent back to the flow solver.

The chemical mechanism used here involves 9 species ($H_2$, $O_2$, OH, $H_2O$, H, O, $HO_2$, $H_2O_2$ and $N_2$) and 21 reactions [42]. At the inlet, the radial profiles of mean velocities and scalars for both jet and coflow are prescribed using the experimental data [11, 14], as tabulated in Table 1; while the turbulent intensity of 5% is used for both the jet and coflow streams. The Neumann boundary condition is applied at the outer radial locations and a symmetry condition is imposed along the centerline of the domain. In the mixture fraction space, the extreme points $\eta=0$ and $\eta=1$ represent the state of pure oxidizer and pure fuel, respectively; whereas at the inlet, linear mixing profiles are used for the conditional mean reactive scalars ($Q_T$, $Q_\alpha$). A convective outflow condition is prescribed at the outlet boundary. The initial flow fields for the RANS-CMC simulations are obtained using the presumed PDF based approach with 'infinite fast chemistry' profiles for temperature and species mass fractions in the mixture fraction space.

## 4. Results and discussion

In this section, we present the numerical predictions using RANS-CMC approach for the lifted jet diffusion flame of hydrogen/nitrogen in a vitiated coflow. Numerical results are compared with the experimental measurements of Cabra *et al.* [11], whenever available. Initially, we discuss the grid independence of the numerical results, followed by a detailed comparison of present predictions for the base case of 1045 K coflow temperature. Finally, the lift-off height variation and flame stabilization are discussed for different parameters, such as conditional velocity models, coflow temperature, mixing model constants, jet- and coflow velocity.

### 4.1. Grid-independence and turbulence model parameters

The numerical simulations are carried out using three different sets of grids: (a) a coarse grid with 72X72 cells, (b) an intermediate grid with 96X96 cells, and (c) a fine grid with 120X120 cells in the radial- and axial directions, respectively using default turbulence parameters. The radial profiles of the Favre mean temperature and the mixture fraction obtained using the coarse, intermediate and fine grids are shown in Figure 2 at $z/D$=8 and 14. The numerical results obtained using different grids are in good agreement with each other and show proper trends of jet spreading and the evolution of mean temperature. Hence, in the subsequent sections, numerical results will be presented for simulations performed using the intermediate grid.

Since, it is very much well known that the standard $k$-$\varepsilon$ turbulence model is not able to capture the jet spreading properly and literatures suggest that it requires some tuning [43-44]. However, the said literatures are primarily based on non-reacting flows and the system becomes completely different while dealing with reacting flows in presence of density and temperature gradient. Masri *et al.* [15] reported that increasing the turbulence constant $C_{\varepsilon1}$ from 1.44 to 1.6 does not reduce the spreading rate of jet and they are of the opinion that such kind of behavior is due to the jet being issued in heated air where the density is less than that of the jet fluid. Furthermore, a similar study carried by Kumar *et al*. [45] revealed the similar behavior. In order to gain more insight, we have also carried out similar studies by varying the turbulence constants and observed that the predictions with default constants appear to be the best one while compared with measurements. Therefore, the rest of the simulations are carried out with intermediate grid with default turbulence constants and reported henceforth.

## 4.2 Effect of velocity closure models

Figure 3 depicts the radial profiles for Favre mean and r.m.s of mixture fraction, Favre mean temperature, $H_2O$ and OH mass fractions for different conditional velocity models, *i.e.*, mean (V1), linear (V2) and gradient (V3) models. Radial profiles of mean mixture fraction are found to be in excellent agreement with the experimental results for all the three models. However, while looking at the variance of mixture fraction, it is observed that the radial profiles of r.m.s of mixture fraction are over-predicted by all the three models around lift-off height region ($z/D$=8). Also, the radial temperature profile is found to be over-predict the experimental results around the lift-off height region ($z/D$=8). The $H_2O$ mass fraction is also found to be over-predicted around the lift-off height region and this behavior can be attributed to the over-prediction of temperature around the same location. The temperature and $H_2O$ mass fraction prediction is found to be in good agreement with experimental results at downstream locations. Sreedhara *et al.* [46] made a systematic comparison between these three models for bluff-body flame and the numerical predictions are reported to be almost similar for all three models, whereas the flow configuration is quite different in the present case, *i.e.*, hot coflow is used as oxidizer. In order to have a better insight about these closure models, the Favre mean and r.m.s axial velocity are plotted in Fig. 4. Since, there is no velocity measurement available, we only present a qualitative assessment of these models. It appears that the liner closure model for the conditional velocity yields better results in terms of the predicted lift-off height. Furthermore, lift-off height predictions using all the three velocity closure models are also analyzed. The lift-off height obtained for all the cases is tabulated in Table 1. The lift-off heights obtained for V1, V2, V3 models are 33%, 68% and 41% of the experimental lift-off height. The accuracy of the gradient model significantly depends upon how well the presumed PDF models imitate the actual PDF, so the results might improve if clipped Gaussian PDF is used instead of $\beta$-PDF. While assessing the performance of all three models, it is quite evitable that the linear velocity closure model predicts the most accurate results and is chosen for the rest of the study.

## 4.3 Validation for the base case

In this section, we present detailed validation results for the base case with a coflow temperature of 1045 K. The simulations are carried out using the linear model for conditional velocity. Radial profiles for the Favre mean temperature, mean and r.m.s. of mixture fraction, mean $H_2O$ and OH mass fractions are plotted in Figure 3 at different axial locations and compared with the measurements [11]. Around the lift-off height, the predicted mean temperature shows over-prediction, which is attributed to the shorter computed lift-off height with respective to the measured lift-off height. The mean mixture fraction is in good agreement with the experimental data at the upstream location, while slight over-prediction in the jet spreading rate is observed from $r/D>1$ at the downstream location. Whereas, the r.m.s. of mixture fraction is found to be consistently over-predicted at all the axial locations and this may be because of the jet issuing into heated co-flow where the density is less than that of the jet fluid. The over-prediction observed in $H_2O$ mass fraction at the upstream location is attributed to the over-prediction of temperature around the same location while it agrees well with the experimental data at downstream locations. Although, the prediction of OH mass fraction agrees reasonably well with experimental data at downstream locations, it is found to be highly over-predicted around the lift-off height location. The lift-off height in this study is measured by identifying the first axial location at any radius, where the OH mass fraction reaches a value of 600 ppm [11]. Since, the predicted lift-off height is found to be less than the experimental lift-off height ($10D$), the variation between predicted and measured OH mass fraction may be considered acceptable. The predictions may be improved by application of the second-order CMC method; however, literature suggests that second order CMC would increase the lift-off height by 10% only [47]. Since these flames are more kinetically controlled, the chemical mechanisms may play an important role and the lift-off height predictions vary a lot [14-16]. Hence, the strong sensitivity of lift-off height to chemical mechanism needs to be investigated in details and can be the scope of future studies.

## 4.4 Effect of co-flow temperature

Figure 5 depicts the radial profiles for mean temperature, mean and r.m.s of mixture fraction for different co-flow temperatures. In addition to the experimentally reported value of 1045K, three other co-flow temperatures are also studied such as: 1010, 1025 and 1035K. Although not much difference is observed in the radial profiles with the change in co-flow temperatures, the lift-off height is found to be very sensitive to the change in coflow temperature and has been discussed in Sec. 4.6.

## 4.5 Conditional results

Figure 6(a) presents the profiles of conditional mean temperature and conditional mass fractions of $H_2O$ and OH at four different axial locations for the base case with a co-flow temperature of 1045K. Around the lift-off height, the peak in the conditional mean temperature profile is observed in the lean mixture. Also, the chemical heat release, and hence conditional mean temperature for the stoichiometric mixture ($\eta_{st}$ =0.47) is found to be smaller than those of the lean mixtures. The peak of conditional mean temperature shifts gradually from lean to the stoichiometric mixture with an increase in the axial distance. In these flames, ignition begins at a downstream location where scalar dissipation rate is low [11, 13, 15] and local mixture corresponding to the "most reactive mixture fraction" [48].

The igniton occuring in the lean parcels can be attributed to the small axial velocity, low scalar dissipation rate and high temperature associated with it. The profile of $Q_T$ around the lift-off height, *i.e.*, *z/D*=7 show that ignition occurs around $\eta$=0.25. This observation can lead to the conclusion that the process is of auto-ignition type at the post flame region. However, elaborate discussion about the flame stabilization process has been made in Sec.4.7. The conditional $H_2O$ mass fraction prediction is found to follow a similar trend as the conditional temperature profile. The conditional OH mass fraction is negligible at the first axial location where pure mixing takes place. The peak OH prediction is found to shift towards $\eta_{st}$ with increasing axial distance. Figures 6(b) – 6(d) depict the profiles of conditional mean temperature and species mass fractions for co-flow temperature 1035, 1025 and 1010K. The observations at these temperatures are also found to be similar. The conditional OH mass fraction profiles are plotted in Fig. 7 at the flame base for different coflow temperature cases. The peak in $\langle Y_{OH}|\eta \rangle$ occurs around $\eta$ = 0.288, 0.265, 0.21, 0.18 for coflow temperature $T_c$ = 1010, 1025, 1035 and 1045 K cases respectively. Mean mixture fraction at the stabilization point of the flame (base of the flame) for all coflow cases remain close to 0.2.

## 4.6 Variation in lift-off height

Numerical simulations are carried in order to determine the variation in lift-off height over a range of parameters, namely, co-flow temperature, mixing model constant $C_\Phi$, jet velocity and coflow velocity for the lifted $H_2/N_2$ jet diffusion flames. The lift-off height is found to be highly sensitive to change in co-flow temperature. Figures 8(a) – 8(c) show the lift-off height variation with co-flow temperature for different values of mixing model constant $C_\Phi$, jet velocity and coflow velocity. The lift-off heights obtained at different coflow temperatures using the standard value of $C_\Phi$=2 have been presented in Table 3. It is observed that the lift-off height decreases with increase in coflow temperature and a decrease of 1% in coflow temperature yields around 20% increase in the predicted lift-off height. This observation can be attributed to the fact that with the increase in coflow temperature, the ignition point moves toward the jet exit plane as the auto-ignition driven stabilization delayed due to lower mixture temperature. The deviation of the lift-off height from the experimental results for the base case could be a consequence of the fact that in the present computation, entrainment of ambient air into hot coflow has not been considered. However, in the experiments, the hot coflow entrains the ambient air resulting in radial non-uniformity in coflow temperature at downstream locations. Further, the standard $k$-$\varepsilon$ model constants also yield a higher jet spreading rate, and hence promote a higher mixing at the flame base. The lift-off height can be improved by changing the value of the mixing model constant $C_\phi$, as tabulated in Table 4. The influence of the mixing model constant on lift-off height is studied with the values of $C_\phi$=1.5, 2.0, 3.0 and 4.0. Figure 8(a) exhibits the variation of lift-off height with increasing $C_\phi$. The difference has been found to be more at low coflow temperatures.

Importantly, it is quite well known, the jet diffusion flames in hot-coflow condition are very much sensitive to the mixing, in turn the ignition; as they are primarily kinetically controlled. Since, the $O_2$ levels are low, the chemical time scale also becomes slower which poses a challenge to model these type of flames accurately. Our primary objective is to assess the impact of $C_\phi$ (turbulent time scale/mixing time scale) on flame stabilization. As observed, for low coflow temperature the higher value of $C_\phi$ yields the best results, because the reaction time scales are much smaller at lower time scale and by slowing down the mixing process between jet & hot coflow, *i.e.* increasing $C_\phi$, the ignition location is nicely captured. Contrarily, lower value of $C_\phi$ provides better results for higher coflow temperature. Hence, there is strong coupling between the coflow temperature (auto-ignition) and the mixing for a particular dilution level. This impact of mixing on these kinetically controlled flames is already been reported in literatures for different hot-coflow flames [49-51]. In the foregoing, we limit ourselves to assess the impact of the mixing constants within the RANS-CMC framework, and hence we do not intend to provide an optimum value of mixing constant for these flames. In light of the present results, one can conclude the optimum value of mixing constant, but that mixing constant will be limited to RANS based calculations only. The present findings are found to be in the similar line as reported by Cao *et al*. [14] in their joint PDF calculations, where they have also observed that the impact of $C_\phi$ is substantial at lower coflow temperatures. It is observed in our study that the results obtained using mixing model constant $C_\phi$=4 appears to be in closest agreement with the experimental results, especially for lower coflow temperatures [12].

Figures 6(a) – 6(d) exhibit that the chemical kinetic activities occur initially in the lean mixture. Hence, the combustion processes of the lifted flame should be more sensitive to coflow conditions rather than jet conditions. This is quite evident from Figs. 8(b) – 8(c), which presents the sensitivity of lift-off height variation to jet velocity and coflow velocity where the lift-off height is found to be more sensitive to coflow velocity mainly at low coflow temperature. The dependence of lift-off height on jet velocity for different temperatures has also been studied for a coflow velocity of 3.5 m/s. Similarly the dependence on coflow velocity for different temperatures is studied for a jet velocity of 107 m/s. The lift-off height is found to increase with increasing values of jet and coflow velocities (Tables 5-6). The linear relationship between the lift-off height and the jet velocity is in agreement with previously reported results of Kalghatgi [3]. The momentum of the coflow stream acts in combination with the jet momentum to push the base of the flame further away. When the gas velocity at the jet exit is higher, it takes longer for the gas velocity to decrease to the point where it matches the propagation rate of the flame and hence shifting the stabilization point further downstream.

## 4.7 Stabilization Mechanism

The CMC equation (Eq.(3)) represents a balance between several terms, namely, the axial and radial mean advection, axial and radial turbulent diffusion, molecular diffusion and chemical source terms in the mixture fraction space. One of the distinct advantages of the CMC method is that it facilitates a thorough investigation of the flame structure in terms of analyzing the conditional fluxes in the mixture fraction space, which can be readily obtained from the computed conditional moments of temperature and species concentrations. In this work, flame stabilization mechanisms of the lifted flames are investigated by plotting different conditional temperature ($Q_T$) flux terms in Eq. (3) along the mixture fraction reference space ($\eta$). Each of these terms, contributing to the equation for $\frac{\partial Q_T}{\partial t}$ is plotted separately in mixture fraction space at the pre-flame and post-flame regions around the lift-off height. A comprehensive analysis of stabilization mechanism for the $H_2/N_2$ lifted diffusion flame will be presented in the following subsections in terms of variation in coflow temperature, mixing model constant, jet and coflow velocities by analyzing the balance between different conditional fluxes, namely, axial convection and diffusion, radial convection and diffusion, molecular diffusion and chemical reaction source terms in the transport equation of conditional moment of temperature. It is important to mention that when analyzing the balance of the conditional fluxes, auto-ignition is the relevant stabilization mechanism if there exists a balance between the axial advection and reaction terms in the pre-flame region, with the axial diffusion term being smaller. Significant radial advection and diffusion fluxes may exist in some cases. On the other hand, if the axial convection flux primarily balances the axial diffusion flux in the pre-flame region, with the chemical reaction term being smaller, then the relevant flame stabilization mechanism is referred as premixed flame propagation.

### 4.7.1 Effect of variation in coflow temperature

Figure 9 shows balance between different conditional flux terms in the mixture fraction space for different coflow temperature cases. The conditional fluxes are presented at two approximate axial locations such that one lies in the preheat zone (upstream of the liftoff height) while other is in the reactive zone (downstream of the liftoff height), whereas the radial location approximately corresponds to that of the flame-base. In the pre-flame region, for all coflow temperature cases, the chemical source term is primarily balanced by the axial advection term, and therefore, autoignition is the relevant flame stabilization mechanism. The peak in the chemical source term decreases with decrease in the coflow temperature and corresponding increase in the lift-off height. Further, the peak in the chemical reaction source term within the mixture fraction space occurs at a lean mixture fraction. Our observation is in accordance with the findings of Gordon *et al.* [16], where they concluded auto-ignition is the flame stabilization mechanism using PDF method. In the reactive zone (post-flame region), the chemistry term is primarily balanced by the molecular diffusion, axial and radial convective terms. The contributions from the axial and radial diffusion terms become negligible. The profiles in this zone are found to be similar for all the co-flow temperatures studied herein. As we move into the reactive zone, the magnitude of chemical source term increases significantly and the peak in the chemical source term is observed close to the stoichiometric mixture fraction ($\eta_{st}$=0.477).

### 4.7.2 Effect of variation in mixing model constant

Shown in Fig. 10 are the balances of various flux terms appearing in the transport equation for conditional temperature using different mixing constants $C_\Phi$ for the base case with the jet velocity $V_{jet}$=107 m/s, coflow temperature $T_c$=1045 K, and the coflow velocity $V_c$=3.5 m/s. At the location upstream of the reaction zone, the chemical source term is dominant especially in the lean mixture and the chemical source term is primarily balanced by the axial advection and molecular source terms. Therefore, it may be concluded that the flame is stabilized by autoignition for all the cases. In the pre-flame region, the peak in the conditional temperature occurs in the lean mixture. Also, the chemical heat release is very weak in the stoichiometric mixture. Depending on the value of the mixing constants $C_\Phi$, the conditional chemical source term profiles change within the pre-flame region. Initially the maximum value of the conditional chemical reaction source term increases up to $C_\Phi$=2.0, and thereafter it decreases with increase in the mixing constant value. Increase in $C_\Phi$ results in more mixing of the scalars (both temperature and species mass fractions) between the jet and coflow streams, which brings in more and more reactants in the reaction zone leading to an increase in the maximum value of the chemical source term when the value of $C_\Phi$ is increased to 2.0. The peak in the chemical source term gradually shifts towards the lean mixture and its peak gradually diminishes with an increase in the value of mixing constant. The molecular diffusion dissipates heat from the reaction zone in the pre-flame region, which results in lower value of the chemical source term. In the post-flame region, for all the cases we observe a classical diffusion flame by the balance between chemistry and molecular diffusion terms.

### 4.7.3 Effect of variation in jet velocity

For all different jet velocity cases investigated here, we observe auto-ignition is the relevant flame stabilization mechanism as in the pre-flame region, primarily the chemical source term is balanced by the axial advection and molecular diffusion terms as shown in Fig. 11. In the pre-flame region, the chemical heat release reaches the maximum for $V_{jet}$ = 107 m/s. However, on the either sides of the jet velocity $V_{jet}$ = 107 m/s ($V_{jet}$= 96 m/s and 120 m/s), the peak in the chemical heat release flux decreases which may be attributed to the increased molecular mixing characterized by the reduction of the molecular diffusion flux term. In the post-flame region, we observe a classical diffusion flame type structure characterized by the balance between the chemical heat release and molecular diffusion terms.

### 4.7.4. Effect of variation in coflow velocity

Shown in Fig. 12 are the balances between different conditional fluxes present in the transport equation for the conditional mean temperature. As the chemical reaction source term is primarily balanced by the axial advection and molecular diffusion terms in the pre-flame region for all coflow cases, the flame stabilization mechanism corresponds to auto-ignition. However, the molecular diffusion increases at the flame base with an increase in the coflow velocity, which results in reduction in the conditional molecular diffusion and chemistry fluxes.

### 5. Conclusion

In the present work, turbulent lifted diffusion flame has been investigated using the first-order CMC method with detailed chemistry. The reaction mechanism used is Mueller mechanism with 9 species and 21 reversible reactions. The $k$-$\varepsilon$ model has been used for solving the turbulent flow field. The results presented in this paper lead to the following conclusions:

1. An evitable difference is observed in the scalar predictions using all three conditional velocity models, however the lift-off height is best predicted using linear model followed by mean and gradient.

2. The predicted flow and scalar fields agree well with the experimental data except in downstream locations of the flame. This might be due to the use of $k$-$\varepsilon$ model which is known to over-predict the spreading of jet.

3. At the lift-off height locations, the conditional mean temperature attains the maximum value on the lean side of stoichiometry. Subsequently, in the downstream location, the most reactive mixture fraction progressively shifts towards the stoichiometric mixture fraction.

4. The flame is highly sensitive to the variation in coflow temperature and a decrease of 1% in the coflow temperature increases the lift-off height by around 20%.

5. A sensitivity study for $C_\Phi$ reveals that the lift-off height improves with an increase in $C_\Phi$ and a value of $C_\Phi=4$ yields the best match with experimental results.

6. The flame lift-off height is more sensitive to the coflow velocity rather than the jet velocity.

7. It is found that autoignition is the relevant flame stabilization mechanism for all the flames investigated here.

**Acknowledgment**

The authors would like to acknowledge the IITK computer center (www.iitk.ac.in/cc) for providing the high-performance computing facility to perform the computations.

.

|  | **Jet** | **Co-flow** |
|---|---|---|
| Velocity (m/s) | 107 | 3.5 |
| Temperature (K) | 305 | 1045 |
| Turbulent kinetic energy (m$^2$/s$^2$) | 1.0 | 0.0726 |
| Dissipation rate (m$^2$/s$^3$) | 1.0 | 19.56 |
| $Y_{H_2}$ | 0.02325 | 0 |
| $Y_{O_2}$ | 0 | 0.1709 |
| $Y_{H_2O}$ | 0 | 0.0645 |
| $Y_{N_2}$ | 0.97675 | 0.7646 |

Table 1: Boundary conditions for the base case as prescribed by Cabra *et al.* [11]. Here *Y* denotes the species mass fraction.

| **Velocity closure models** | **Lift-off height (*h/D*)** |
|---|---|
| V1 | 4.2 |
| V2 | 6.8 |
| V3 | 3.3 |

Table 2: Lift-off heights (*h/D*) for the base case obtained using different velocity closure models

| **Coflow temperature (K)** | **Lift-off height (*h/D*)** |
|---|---|
| 1010 | 13.27 |
| 1025 | 10.61 |
| 1035 | 8.84 |
| 1045 | 6.8 |

Table 3: Lift-off heights (*h/D*) for different coflow temperature ($V_{jet}$=107m/s, $V_c$=3.5m/s, $C_\Phi$=2)

| ***Coflow temperature(K)*** →  | | | | |
|---|---|---|---|---|---|
| ***C$_\Phi$*** | | **1010** | **1025** | **1035** | **1045** |
| ↓ | **1.5** | 10.45 | 8.44 | 7.64 | 5.73 |
| | **2.0** | 13.27 | 10.61 | 8.84 | 6.8 |
| | **3.0** | 19.17 | 13.35 | 11.21 | 8.0 |
| | **4.0** | 29.8 | 15.55 | 11.85 | 9.02 |

Table 4: Lift-off heights (*h/D*) for different coflow temperature and *C$_\Phi$*

| *Coflow temperature (K)* | | | | | |
|---|---|---|---|---|---|
| *V*$_c$ | | **1010** | **1025** | **1035** | **1045** |
| *(m/s)* | **3.5** | 13.27 | 10.61 | 8.84 | 6.8 |
| | **7.0** | 21.64 | 13.81 | 12.23 | 7.99 |

Table 5: Lift-off heights (*h/D*) for different coflow velocity (*V*$_{jet}$=107m/s, *C$_\Phi$*=2)

| *Coflow temperature (K)* | | | | | |
|---|---|---|---|---|---|
| *V*$_{jet}$ | | **1010** | **1025** | **1035** | **1045** |
| **(m/s)** | **96** | 12.20 | 9.42 | 7.48 | 5.10 |
| | **107** | 13.27 | 10.61 | 8.84 | 6.8 |
| | **120** | 15.51 | 11.94 | 9.72 | 8.2 |

Table 6: Lift-off heights (*h/D*) for different jet velocity (*V*$_c$=3.5m/s, *C$_\Phi$*=2)

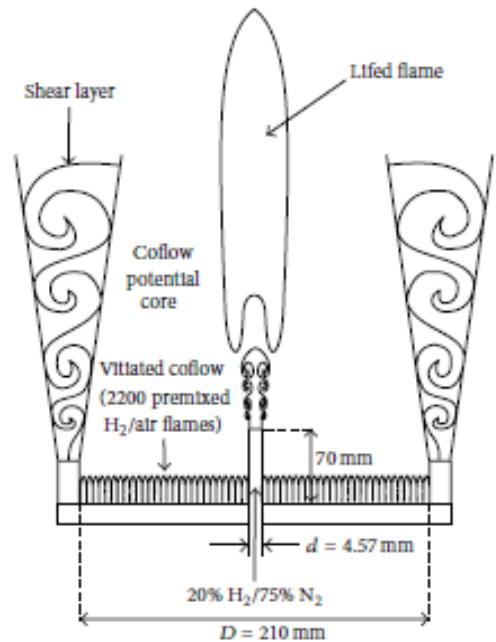

Figure 1: Schematic diagram of lifted jet flame in vitiated co-flow [11]

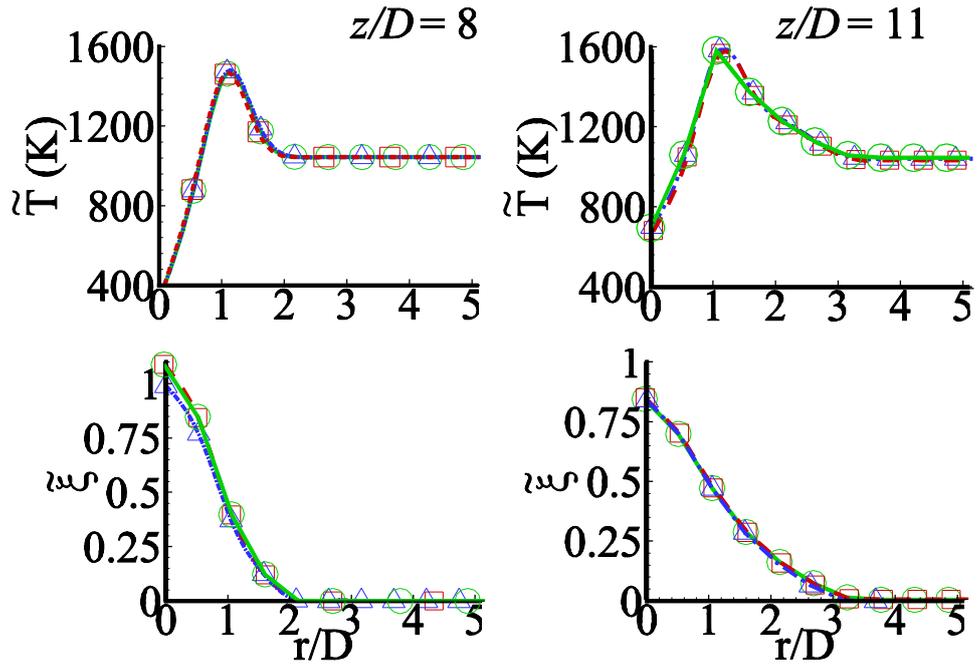

Figure 2: Radial profiles of the Favre mean temperature and the mixture fraction for coflow temperature of 1045K at axial locations $z/D$=8 and 14 (solid lines with circles: fine grid; dashed lines with squares: intermediate; dashed-dot lines with triangle: coarse grid).

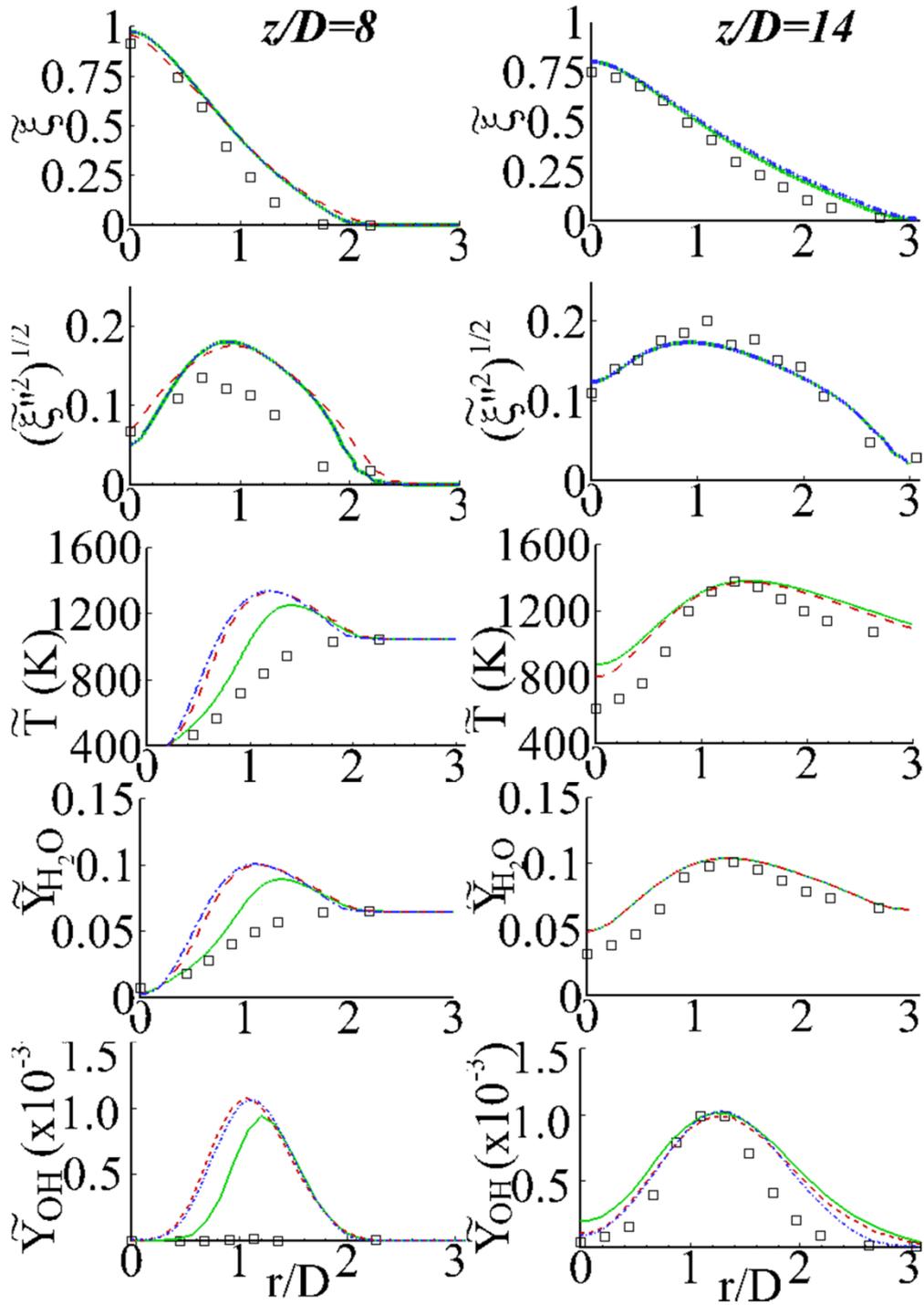

Figure 3: Radial profiles of the Favre mean and r.m.s of mixture fraction, Favre mean temperature, Favre mean mass fractions of H$_2$O and OH at axial locations $z/D$=8 and 14 for coflow temperature of 1045K obtained using different models for the conditional velocity (solid lines: linear; dashed lines: mean; dashed-dotted lines: gradient; symbols: experimental data of Cabra *et al.* [11].

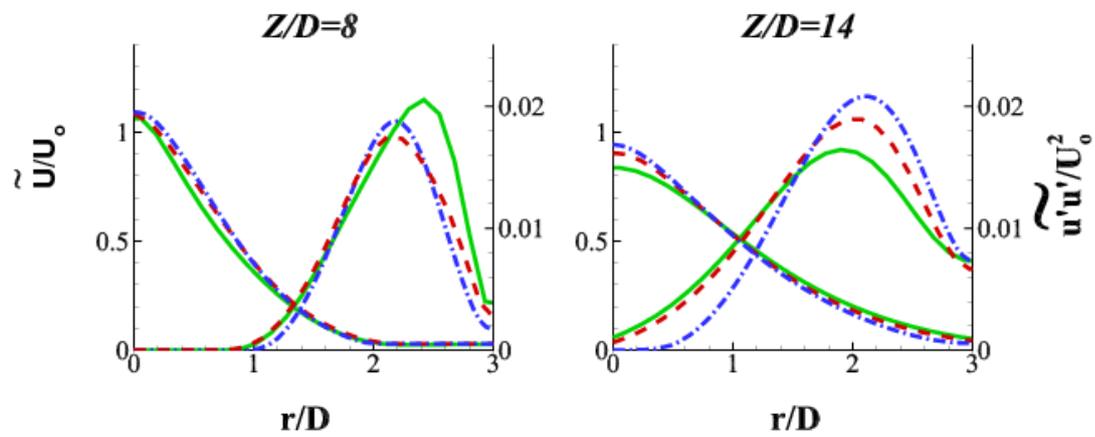

Figure 4: Radial profiles of the Favre mean and r.m.s of axial velocity at axial locations *z/D*=8 and 14 for coflow temperature of 1045K obtained using different models for the conditional velocity (solid lines: linear; dashed lines: mean; dashed-dotted lines: gradient)

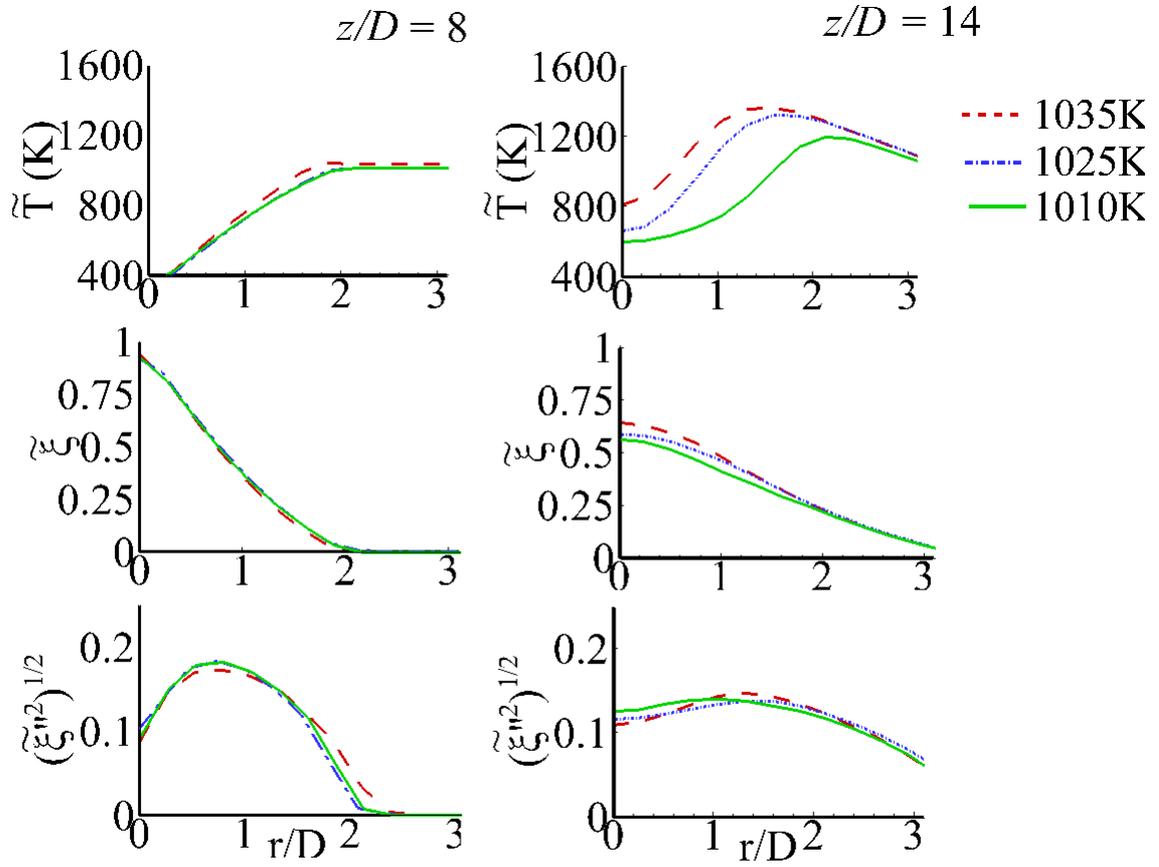

Figure 5: Radial profiles of the mean temperature, mixture fraction and r.m.s of mixture fraction for different co-flow temperature using linear conditional velocity model.

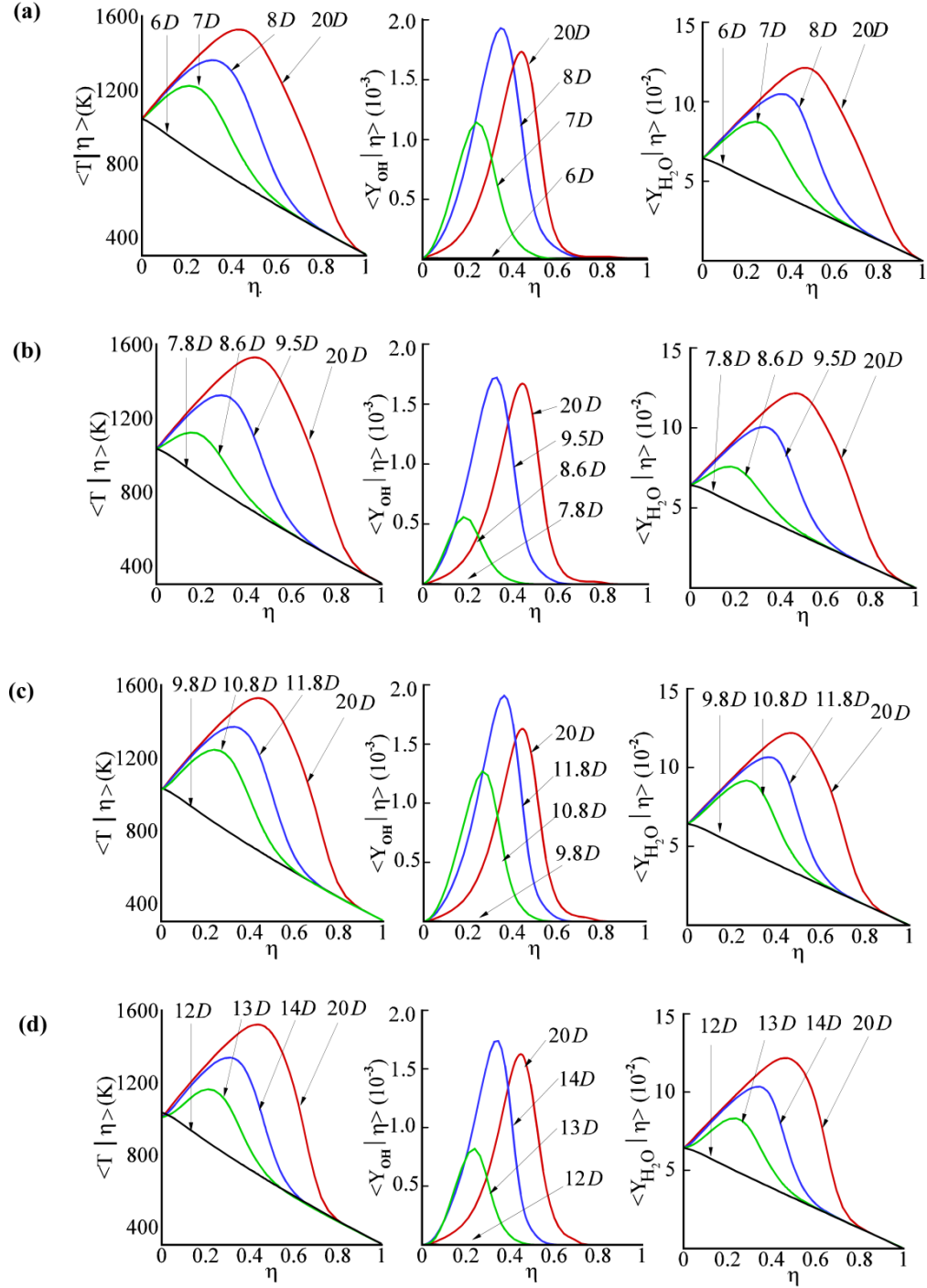

Figure 6: Profiles of conditional mean temperature, OH and $H_2O$ mass fractions in mixture fraction space at different axial locations: (a) at $r/D$=1.32 for a co-flow temperature of 1045K, (b) at $r/D$=1.52 for a co-flow temperature of 1035K, (c) at $r/D$=1.72 for a co-flow temperature of 1025K, (d) at $r/D$=1.93 for a co-flow temperature of 1010K.

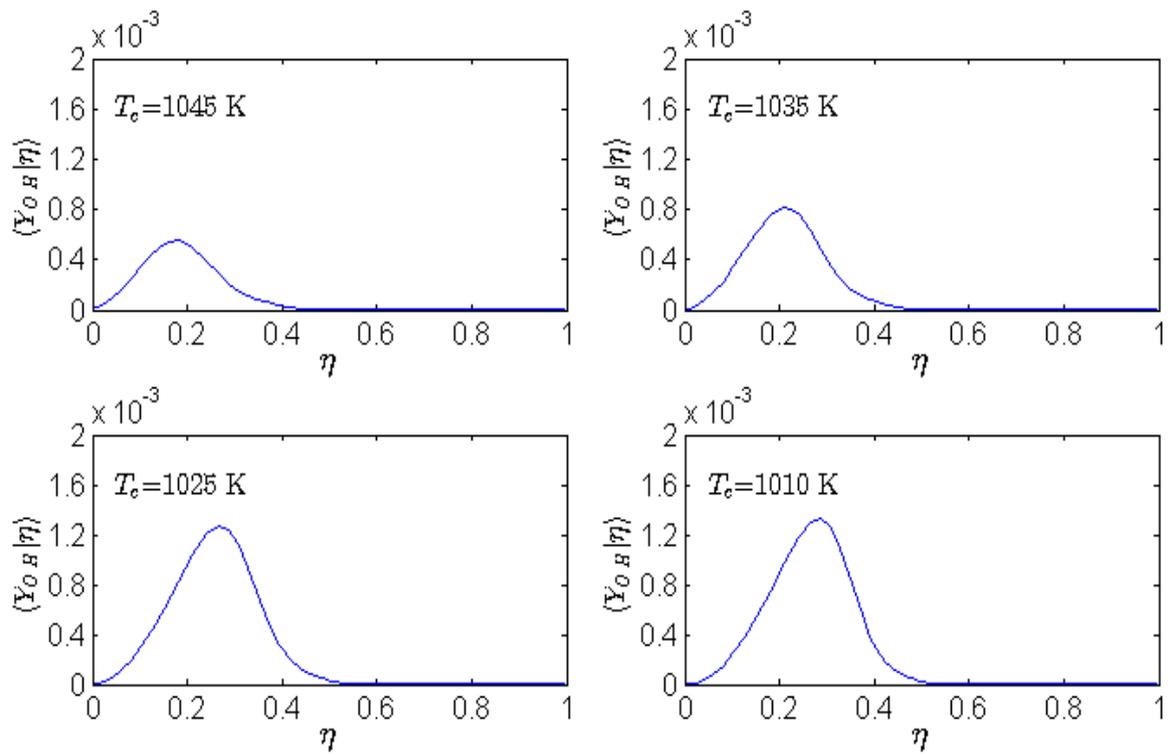

Figure 7: Profiles of conditional OH mass fraction in mixture fraction space at the flame base for various co-flow temperature cases.

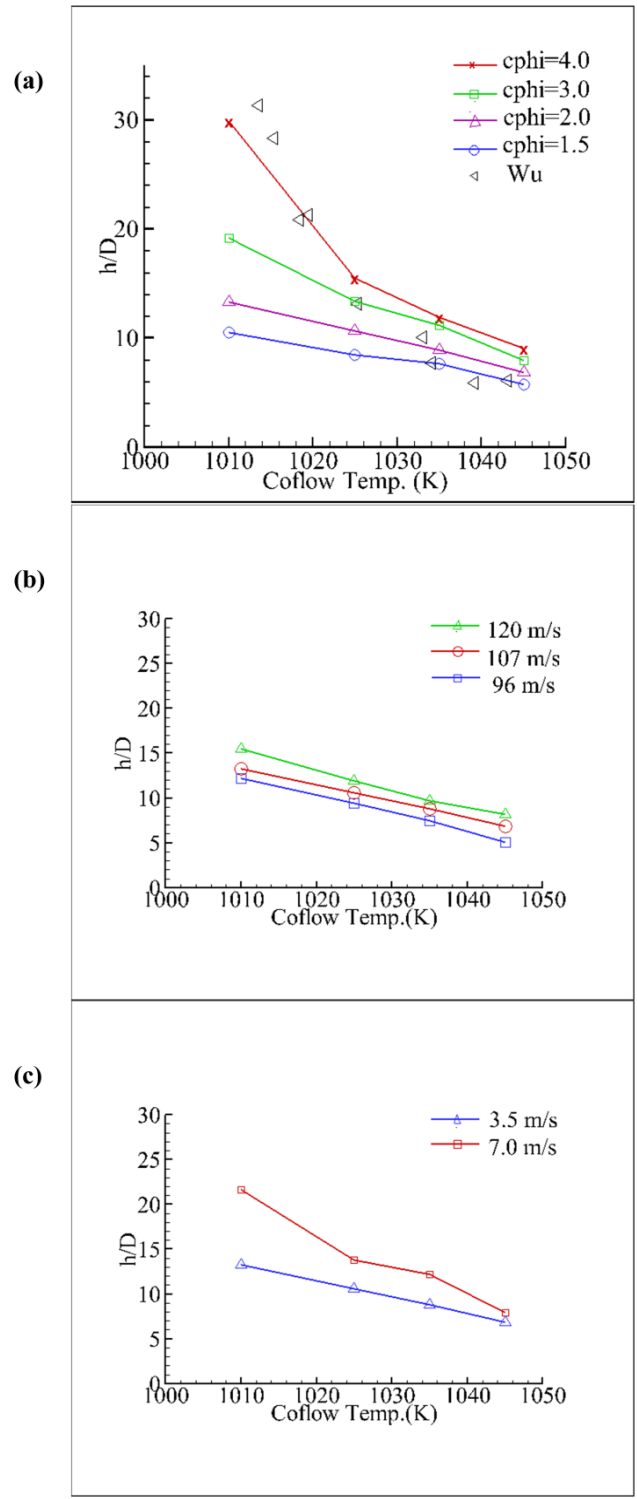

Figure 8: Variation in lift-off height against coflow temperature: (a) effect of different mixing model constant, (b) effect of jet velocities, and (c) effect of coflow velocities.

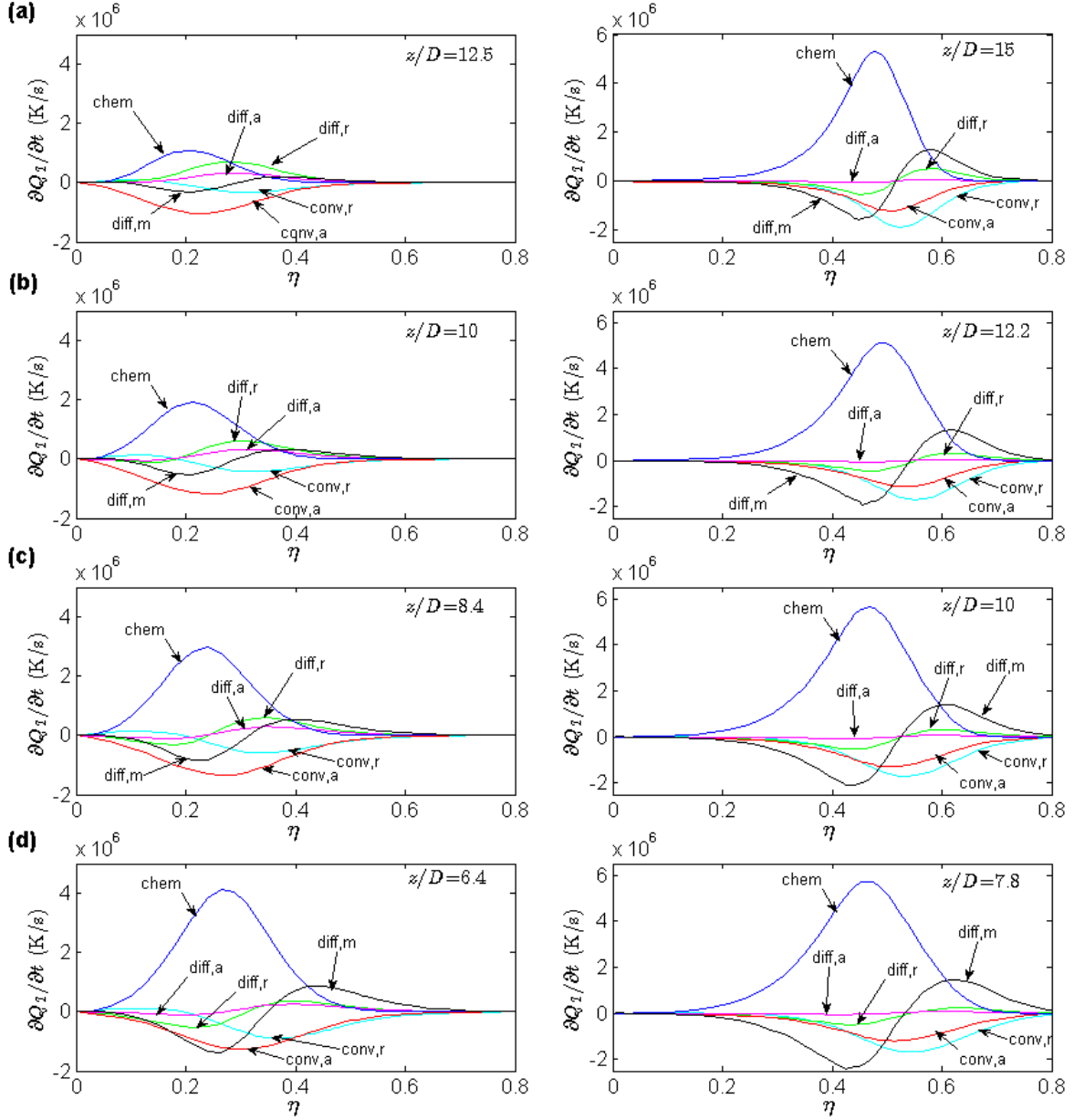

Figure 9: Different terms contributing to $\partial Q_T / \partial t$, in the mixture fraction space, in the pre-flame (left column) and flame (right column) regions, for the cases (a) $T_c = 1010$ K at $r/D$=1.93, (b) $T_c = 1025$ K at $r/D$=1.72, and (c) $T_c = 1035$ K at $r/D$=1.52, and (d) $T_c = 1045$ K at $r/D$=1.32. The jet velocity $V_{jet}$=107 m/s and the coflow velocity $V_c$=3.5 m/s are same for all the cases (diff,a: axial diffusion; diff,r: radial diffusion; conv,a: axial convection; conv,r: radial convection; diff,m: molecular diffusion; chem: chemical source term).

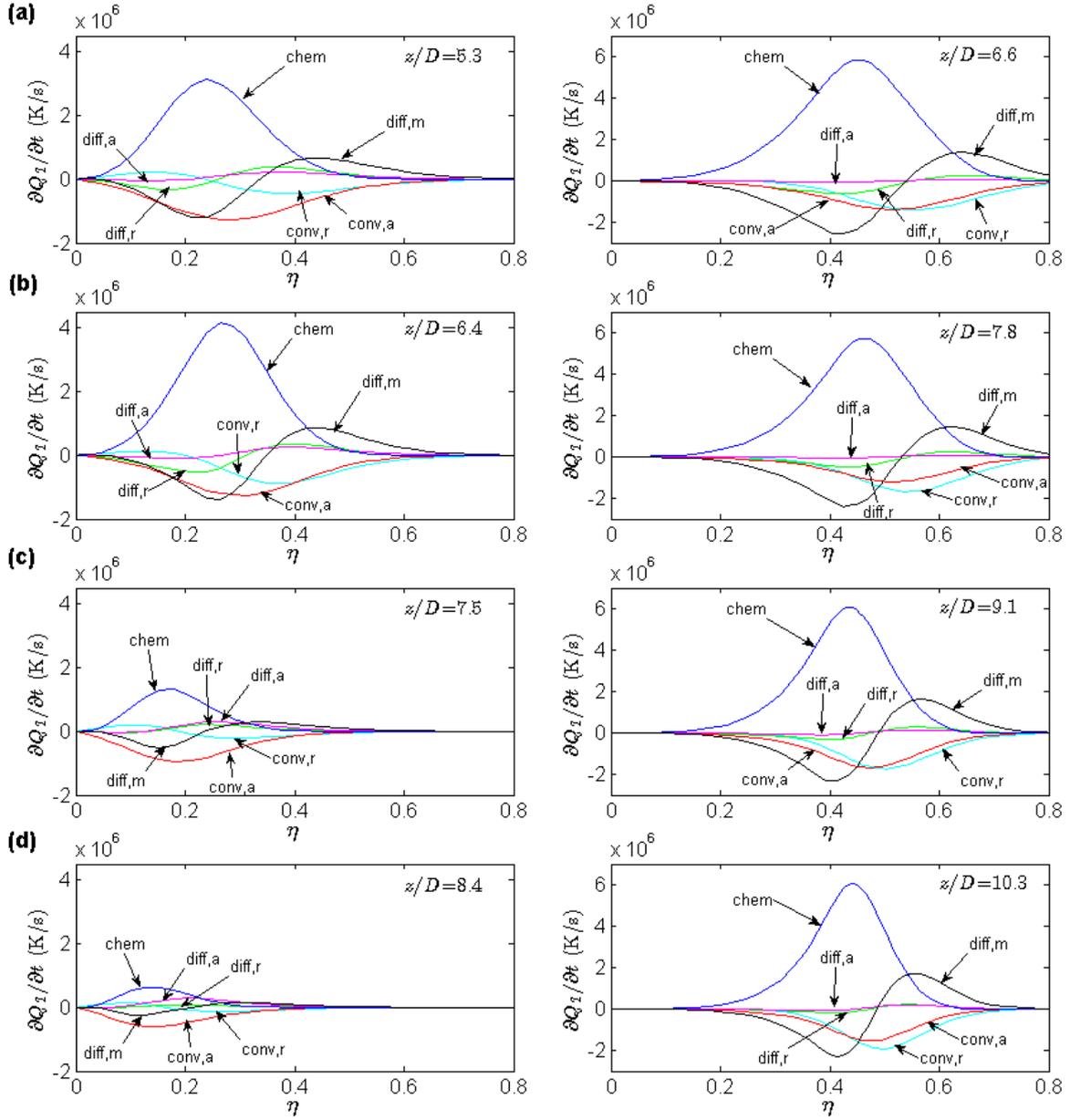

Figure 10: Different terms contributing to $\partial Q_T/\partial t$, in the mixture fraction space, in the pre-flame (left column) and flame (right column) regions, for the cases (a) $C_\varphi$=1.5 at $r/D$=1.16, (b) $C_\varphi$=2 at $r/D$=1.32, and (c) $C_\varphi$ =3 at $r/D$=1.43, and (d) $C_\varphi$=4 at $r/D$=1.6. The jet velocity $V_{jet}$=107 m/s, coflow velocity $T_c$=1045 K, and the coflow velocity $V_c$=3.5 m/s are same for all the cases (diff,a: axial diffusion; diff,r: radial diffusion; conv,a: axial convection; conv,r: radial convection; diff,m: molecular diffusion; chem: chemical source term).

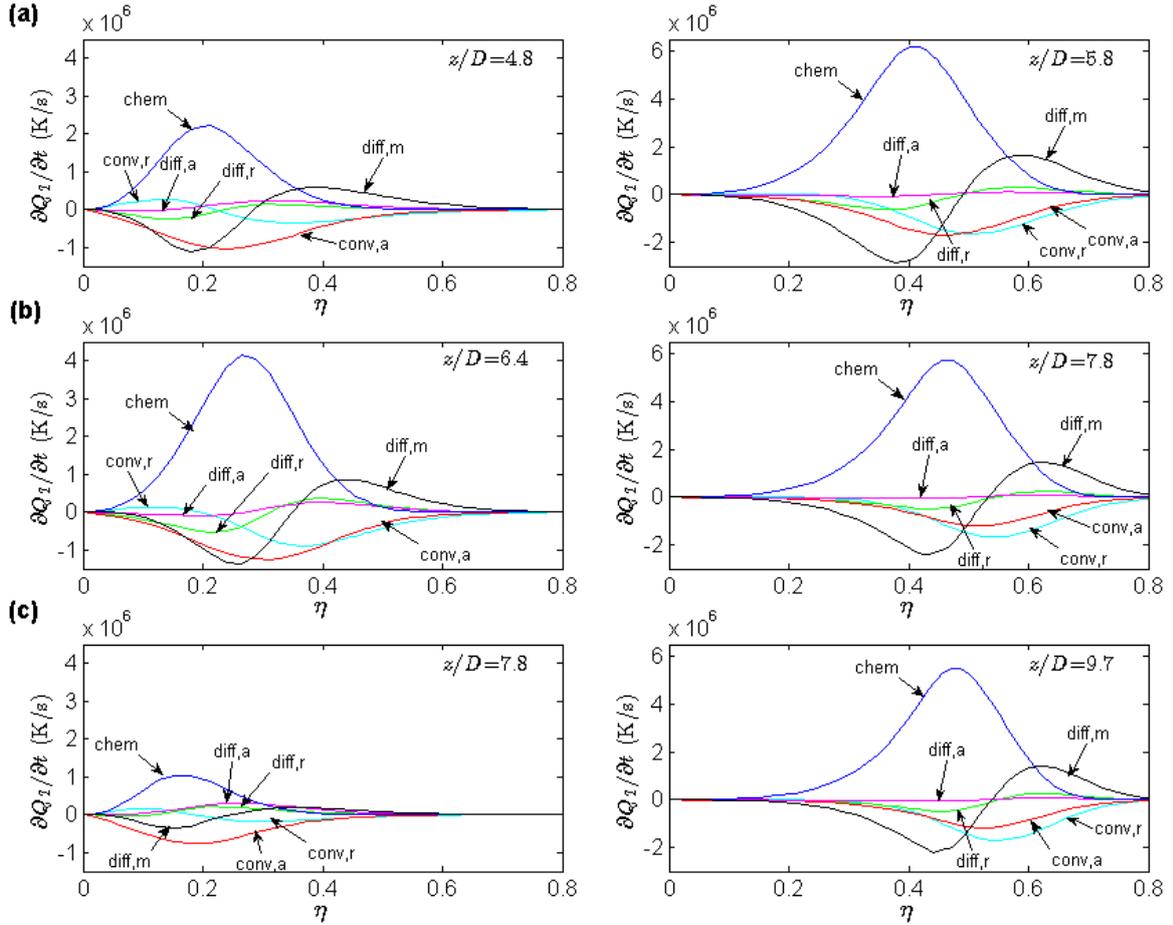

Figure 11: Different terms contributing to $\partial Q_T/\partial t$, in the mixture fraction space, in the pre-flame (left column) and flame (right column) regions, for the cases (a) $V_{jet}$ = 96 m/s at $r/D$=1.1 (top row), (b) $V_{jet}$ = 107.0 m/s at $r/D$=1.32 (middle row), and (c) $V_{jet}$ = 120 m/s at $r/D$=1.52 (bottom row). The coflow temperature $T_c$=1045 K and the coflow velocity $V_c$=3.5 m/s are same for all the cases (diff,a: axial diffusion; diff,r: radial diffusion; conv,a: axial convection; conv,r: radial convection; diff,m: molecular diffusion; chem: chemical source term).

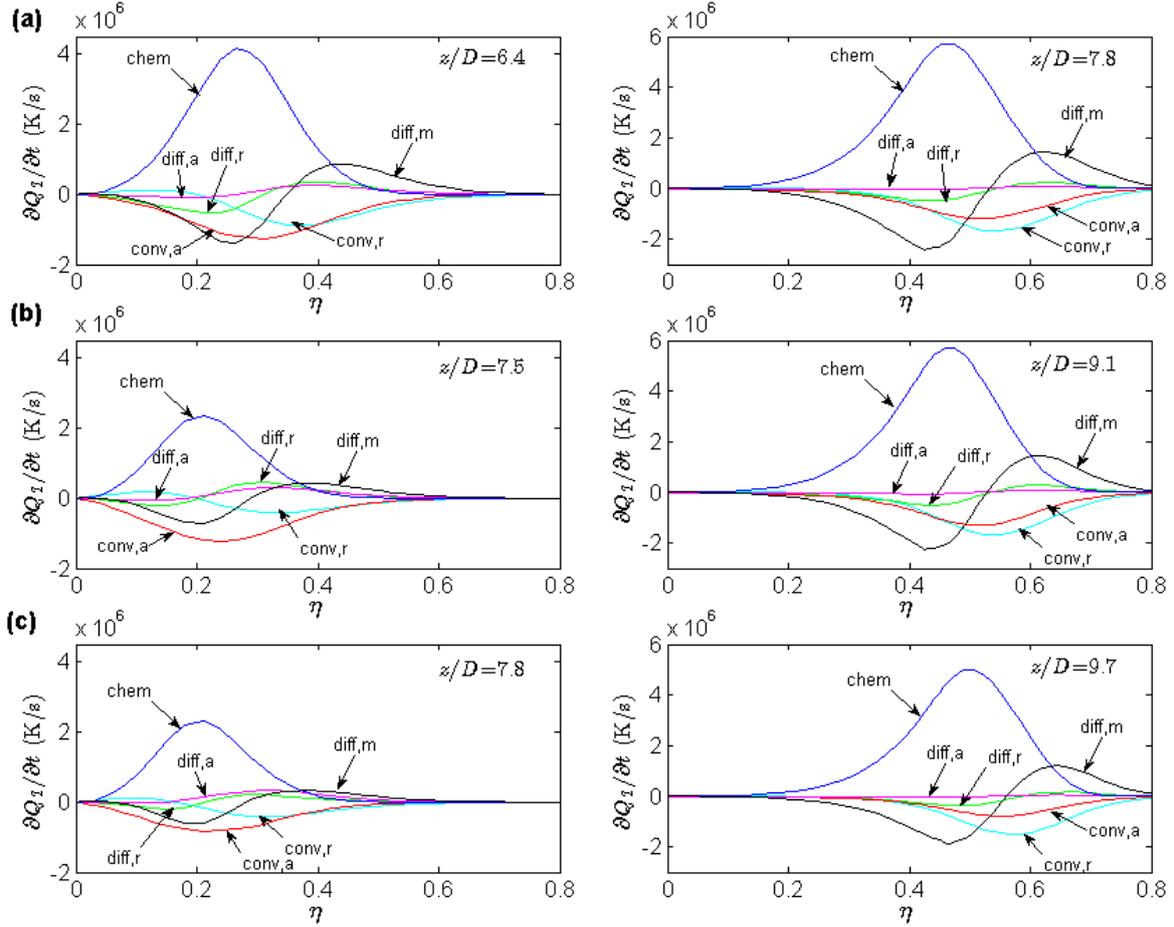

Figure 12: Different terms contributing to $\partial Q_T / \partial t$, in the mixture fraction space, in the pre-flame (left column) and flame (right column) regions, for the cases (a) $V_c$ = 3.5 m/s at $r/D$=1.32 (top row), (b) $V_c$ = 7.0 m/s at $r/D$=1.43 (middle row), and (c) $V_c$ = 10.5 m/s at $r/D$=1.52 (bottom row). The coflow temperature $T_c$=1045 K and the jet velocity $V_{jet}$=107 m/s are same for all the cases (diff,a: axial diffusion; diff,r: radial diffusion; conv,a: axial convection; conv,r: radial convection; diff,m: molecular diffusion; chem: chemical source term).